\documentclass{aa}  
\usepackage[varg]{txfonts}
\usepackage{graphicx}
\usepackage{amssymb}
\usepackage{pifont}
\usepackage{soul}
\newcommand\numberthis{\addtocounter{equation}{1}\tag{\theequation}} 
\usepackage{txfonts}
\usepackage{hyperref}
\usepackage[none]{hyphenat}
\begin{document} 

\renewcommand{\arraystretch}{0.5}

\title{The OTELO survey}
\subtitle{II. The faint-end of the H$\alpha$ luminosity function at ${\sf z}\sim0.40$}
\author{Marina Ram\'on-P\'erez\inst{1,2}
\and \'Angel Bongiovanni\inst{1,2,3}
\and Ana Mar\'ia P\'erez Garc\'ia\inst{4,3}
\and Jordi Cepa\inst{1,2,3}
\and Maritza A. Lara-L\'opez\inst{6}
\and Jos\'e A. de Diego\inst{7}
\and Emilio Alfaro\inst{8}
\and H\'ector O. Casta\~neda\inst{9}
\and Miguel Cervi\~no\inst{1,8,4}
\and Mirian Fern\'andez-Lorenzo\inst{8}
\and Jes\'us Gallego\inst{10}
\and J. Jes\'us Gonz\'alez\inst{7}
\and J. Ignacio Gonz\'alez-Serrano\inst{11,3}
\and Jakub Nadolny\inst{1,2}
\and Iv\'an Oteo G\'omez\inst{12,13}
\and Ricardo P\'erez Mart\'inez\inst{5,3}
\and Irene Pintos-Castro\inst{14}
\and Mirjana Povi\'c\,\inst{8,15}
\and Miguel S\'anchez-Portal\inst{16,17,3}
}

\institute{Instituto de Astrof\'isica de Canarias (IAC), E-38200 La Laguna, Tenerife, Spain 
\and Departamento de Astrof\'isica, Universidad de La Laguna (ULL), E-38205 La Laguna, Tenerife, Spain 
\and Asociaci\'on Astrof\'isica para la Promoci\'on de la Investigaci\'on, Instrumentaci\'on y su Desarrollo, ASPID, E-38205 La Laguna, Tenerife, Spain 
\and Centro de Astrobiolog\'ia (CSIC/INTA), 28850 Torrej\'on de Ardoz, Madrid, Spain 
\and ISDEFE for European Space Astronomy Centre (ESAC)/ESA, P.O. Box 78, E-28690, Villanueva de la Ca\~nada, Madrid, Spain 
\and Dark Cosmology Centre, Niels Bohr Institute, University of Copenhagen, Juliane Maries Vej 30, 2100 Copenhagen {\O}, Denmark %
6
\and Instituto de Astronom\'ia, Universidad Nacional Aut\'onoma de M\'exico, M\'exico D.F., Mexico 
\and Instituto de Astrof\'isica de Andaluc\'ia, CSIC, E-18080, Granada, Spain  
\and Departamento de F\'isica, Escuela Superior de F\'isica y Matem\'aticas, Instituto Polit\'ecnico Nacional, M\'exico D.F., Mexico 
\and Departamento de F\'isica de la Tierra y Astrof\'isica \& Instituto de F\'isica de Part\'iculas y del Cosmos (IPARCOS), Facultad de CC. F\'isicas, Universidad Complutense de Madrid, E-28040, Madrid, Spain 
\and Instituto de F\'isica de Cantabria (CSIC-Universidad de Cantabria), E-39005 Santander, Spain 
\and Institute for Astronomy, University of Edinburgh, Royal Observatory, Blackford Hill, Edinburgh, EH9  3HJ, UK 
\and European Southern Observatory, Karl-Schwarzschild-Str. 2, 85748, Garching, Germany  
\and Department of Astronomy \& Astrophysics, University of Toronto, Canada 
\and Ethiopian Space Science and Technology Institute (ESSTI), Entoto Observatory and Research Center (EORC), Astronomy and Astrophysics Research Division, PO Box 33679, Addis Abbaba, Ethiopia 
\and European Southern Observatory, Alonso de C\'ordova 3107, Vitacura Casilla 19001, Santiago, Chile  
\and Joint ALMA Observatory, Alonso de C\'ordova 3107, Vitacura Casilla 763 0355, Santiago, Chile 
}
\date{Received 24 April 2018 / Accepted 30 November 2018}

  \abstract
   {}
   {We take advantage of the capability of the OTELO survey to obtain the H$\alpha$ luminosity function (LF) at ${\rm z}\sim0.40$. Because of the deepest coverage of OTELO, we are able to determine the faint end of the LF, and thus better constrain the star formation rate and the number of galaxies at low luminosities. The AGN contribution to this LF is estimated as well.}
   {We make use of the multiwavelength catalogue of objects in the field compiled by the OTELO survey, which is unique in terms of minimum flux and equivalent width. We also take advantage of the pseudo-spectra built for each source, which allow the identification of emission lines and the discrimination of different types of objects.}
   {The H$\alpha$ luminosity function at $z\sim0.40$ is obtained, which extends the current faint end by almost 1 dex, reaching minimal luminosities of $\log_{10}L_{\rm lim}=38.5$ erg s$^{-1}$ (or $\sim0.002\, \text{M}_\odot\text{ yr}^{-1})$. The AGN contribution to the total H$\alpha$ luminosity is estimated. We find that no AGN should be expected below a luminosity of $\log_{10}L=38.6$ erg s$^{-1}$. From the sample of non-AGN (presumably, pure SFG) at $z\sim0.40$ we estimated a star formation rate density of  $\rho_{\rm SFR}=0.012\pm0.005\ {\rm \text{M}_{\odot}\ yr^{-1}\ Mpc^{-3}}$.}
   {}

   \titlerunning{The LF(H$\alpha$) at z$\sim$0.4 in the OTELO survey}
   \authorrunning{Ram\'on-P\'erez et al.}
   \maketitle
%

\section{Introduction}

The luminosity function (LF) is an essential empirical tool to evaluate the distribution and large-scale structures of galaxies in the Universe. This function gives the number density $\Phi$ (erg s$^{-1}$ Mpc$^{-3}$) of galaxies per luminosity interval. By tracing specific emission lines across different redshifts, the evolution of star-forming galaxies can also be studied. Given its importance, the LF is usually one of the first things to be derived in any survey. However, this is not always a simple task, as corrections from incompleteness and extinction, among other, must be made (see \citealt{Johnston2011} for a review on the topic).\par

In the case of the H$\alpha$ emission line, the LF allows us to estimate the star formation rate (SFR) function over different cosmological times (see \citealt{Gallego1995}), giving invaluable information about the way our Universe has evolved. \cite{Sobral2013} studied the evolution of the H$\alpha$ LF between redshifts $z$ = 0.40 and 2.23, emphasising the high sensitivity of this SFR tracer by far when compared to widely used proxies in the ultraviolet (UV) or the far-infrared (FIR) domain. Even so, they claim that the evolution seen in H$\alpha$ LF in the last 11 Gyr is in agreement with those obtained using UV and FIR tracers. A similar agreement (within errors) among these SFR functions have been previously reported by for example \cite{Martin2005} and \cite{Bothwell2011} in the case of local Universe, who sampled some decades in luminosity up to lower limits of $\log_{10} L{\rm (UV; FIR)}= 6-7\, L_{\odot}$. A similar trend is reported by \cite{Magnelli2009} from measuring the SFR history at $0.4 < z < 1.4$ ($11 < \log_{10} L_{\rm IR} < 13\, L_{\odot}$), and by \cite{Smit2012} at $z > 4$. However, we identified a significant discrepancy between the results given by \cite{Sobral2013} at low redshift and those of \cite{Drake2013} or \citealt{Ly2007}, especially at low luminosities, which corresponds to the faint end of the SFR function. Indeed, there is evidence of the strong dependence of the faint-end slope on the environment at selected redshifts, even within the framework of large extragalactic surveys (e.g. \citealt{Sobral2011, Geach2012}).

The faint end of the SFR distribution functions allows us to quantify the contribution of low-mass galaxies with mild star formation, which are always more numerous than starburst (i.e. SFR $>$ 10 M$_{\odot}$ yr$^{-1}$) galaxies, to the SFR density estimation at a given epoch. It also provides clues about the processes involved in galaxy formation processes at small dark matter (DM) halo scales and the feedback effects on the star formation of low-luminosity galaxies \citep{Bothwell2011}, in contrast with model predictions (i.e. the dubbed ``missing satellite'' problem, \citealt{Klypin1999, Moore1999}). 
As a consequence, obtaining deep spectroscopic data to extend the statistics of star-forming galaxies towards very faint luminosities provides vital insights to constrain the SFR functions and unravel the causes behind the inconsistencies cited above.\par 

Complementary to this, there is growing evidence of the role of active galactic nuclei (AGN) in the regulation of star formation processes in their hosts \citep{Azadi2015}. In this sense, the AGN contamination must be taken into account in order to construct the SFR function. Even though the AGN contribution seems to be increased with the stellar mass (e.g. \citealt{Xue2010}) for a given redshift, the fraction of these objects is unevenly estimated depending on the characteristics of the survey. For instance, \cite{Garn2010} claimed that the overall AGN contamination in their sample is between 5 and 11\%, but that these numbers probably underestimate the real rate. Conversely, \cite{Sobral2016} found that 30\% of their objects are AGN and that this fraction increases strongly with H$\alpha$ luminosity;  \cite{Matthee2017} also reached the same conclusion. If the AGN population at $z\sim0.40$ can be figured out, the contribution of active galaxies to the LF can be directly derived without the need of additional assumptions or estimations, as in most works. \par 

In this work, we make use of data from OTELO, an ambitious narrow-band survey using the red tunable filter of the 10.4 m Gran Telescopio Canarias (La Palma, Spain). The OTELO survey targets a $7.5\ \times\ 7.4$ arcmin$^2$ region of the Extended Groth Strip (EGS) with the aim of detecting emission-line sources. To this purpose, a tomography of 36 scans with a full width at half maximum (FWHM) of 12 \AA$ $ and a sampling interval of 6 \AA$ $ is made in the range 9070-9280 \AA$ $, allowing the construction of a pseudo-spectra for each source in the field. Details on the OTELO survey, the use of tunable filters and its data products are provided in \cite{OTELO1}.
\par 

With these characteristics, OTELO has become the deepest emission-line survey to date, unique in terms of minimum flux and equivalent width (EW). In particular, the H$\alpha$+[NII] lines are observed in OTELO at $z\sim0.40$. A multiwavelength catalogue of all the sources detected in the field, with data ranging from X-rays to FIR, has already been compiled. The OTELO multiwavelength catalogue, together with the pseudo-spectra, allow for the identification of the H$\alpha$ emitters in the field at $z\sim 0.40$ (both AGN and non-AGN), enabling the construction of the corresponding segregated LF.\par

We therefore aim to take advantage of the capability of OTELO to obtain the H$\alpha$ LF to extend its faint end and thus constraining the SFR and the number of galaxies at low luminosities. Exploiting the AGN selection and analysis in OTELO survey made by \cite{ramonperez19}, hereafter referred to as OTELO-III, and carrying out the pertaining diagnostics, the AGN contribution to this LF is also estimated.\par 

This paper is organised as follows. In section \ref{selectionidentificationhalpha}, the selection and identification of  the sample of H$\alpha$ emitters in OTELO is explained, as well as the H$\alpha$ line flux measurements and the detection limits in OTELO for this particular line. The construction of the LF is then described in section \ref{LF_sec}. Finally, section \ref{conclusions} summarises the main conclusions of this work.\par
Through this paper a standard cosmology with $\Omega_\Lambda$=0.7, $\Omega_{\rm m}$= 0.3, and $H_0$=70 km s$^{-1}$ Mpc$^{-1}$ is assumed for consistency with recent contributions related to the main topic.

\section{Identification of H$\alpha$ emitters, line flux measurements, and detection limits} 
\label{selectionidentificationhalpha}

Emission-line objects in OTELO were first selected using both an automatic algorithm and a visual classification, based on the appearance of emission significant features in the pseudo-spectra from OTELO. Then, their photometric redshifts were obtained using {\it LePhare} code (\citealt{Arnouts1999}, \citealt{Ilbert2006}) and ancillary data from OTELO. A complementary search for emitting candidates was performed by looking for narrow-band excess and the location of OTELO sources in a colour-colour diagram. A detailed description of the first selection of line emitters in OTELO survey is described in \cite{OTELO1}. We focus on the selection of H$\alpha$ line emitters and the associated detections biases.

\subsection{Selection of H$\alpha$ emitters candidates}

Following the methodology described in \citealt{OTELO1}, a sample of potential H$\alpha$ emitters candidates in OTELO was first obtained. It was composed of all the objects selected by one or more of the different methods as follows: by their spectroscopic or photometric redshifts (4 and 108 objects, respectively), location in the colour-colour diagram (32 objects), or narrow-band excess (27 objects). \par 
This selection was deliberately broad enough to ensure a sample as complete as possible of the existing H$\alpha$ emitters in the OTELO catalogue. A redshift range of $0.30\leq z \leq0.50$ was chosen for the search of potential H$\alpha$ candidates to take into account the dispersion of the photometric redshift estimations, whose accuracy is better than $\vert\Delta$z$\vert$/(1+z) $\leq$ 0.2 (\citealt{OTELO1}). Additionally, 31 emitters that did not fulfil any of the previously mentioned conditions, but whose pseudo-spectra show signs of a possible H$\alpha$+[NII] emission, were also included in the sample. In total, the final sample of candidates comprised 202 objects, some of which were selected by more than one of the mentioned criteria. Table \ref{halpha_candidates} lists a summary of the number of H$\alpha$ emitters candidates selected by each criterion.

\begin{table}[ht]
\vspace*{4mm} 
\caption[Selection of H$\alpha$ candidates sample]{Selection of H$\alpha$ candidates sample. Column 1: criterion used for the selection of the candidate. Column 2: number of candidates selected with that criterion. Some objects may have been selected by more than one criterion but they are only included in one category, following the order of the list from top to bottom.}       
\vspace*{-3mm}       
\label{halpha_candidates}      
\centering 
  \small                                 
\begin{center}\begin{tabular}{c c c}          
\hline   \\   [-1pt]                   
   Criterion & $N$    \\  
\hline      \\[3pt]                        
     with $0.37\leq z_{\rm spec}\leq 0.42$ & 4   \\[4pt] 
    with $0.30\leq z_{\rm phot}\leq 0.50$ & 108   \\[4pt] 
     selected by colour &  32    \\[4pt] 
  selected by narrow-band excess & 27      \\[4pt] 
selected by appearance & 31 \\[6pt]
\textbf{TOTAL} &  \textbf{202} \\
\hline                                             
\end{tabular}\end{center}
\end{table}
\normalsize

\subsection{Identification of H$\alpha$ emitters}
\label{sec_identification}

To ensure reliable data, each object in the sample of H$\alpha$ candidates was examined by four independent collaborators. These collaborators used a web-based graphical user interface (GUI) specifically designed as an on-line platform to have an up-to-date dossier of any OTELO source, and as a facility for the analysis of selected objects, in particular the labelling of significant emission/absorption features in the pseudo-spectrum. For this task the GUI includes a line identifier tool that contains an extensive list (available in OTELO survey URL\footnote{\tt http://research.iac.es/proyecto/otelo}) of relevant traits that would be seen in the pseudo-spectrum at a selected redshift (including the relative strengths depending on the emission-line object type). The GUI includes three main sections: 1) General Analysis, which contains the source identification and the information about the properties of the emission line(s) in its pseudo-spectrum, 2) Image Analysis, where the postage-stamp images of the source in the ancillary photometric bands are analysed to search for possible blended/multiple sources or dropouts and obtain clues about their morphology, and 3) Photo-z Analysis, where the information about spectral energy distribution fitting performed with {\it LePhare} code, including uncertainties, and the secondary redshift solutions if available, are taken into account to provide a redshift guess that is refined when the line flux is measured. The web-based GUI was designed to accept prepared lists of sources and store value-added data provided by the collaborators after the source analysis, as demonstrated below, but it also can be used as a free browser for visual inspection of individually selected objects.\par

Considering all the information gathered in the GUI, each one of the 202 H$\alpha$ candidates was analysed and assigned up to three possible values of redshift by each collaborator, with a corresponding likelihood value or degree of confidence. This likelihood is scaled from $L=5$ (highly reliable redshift) to $L=1$ (possible but not very reliable).
 The probabilities of the redshift of an object to belong (or not) to the OTELO H$\alpha$ window ($0.37 \leq z \leq 0.42$) were then calculated by comparing and weighting the different values of redshift and the corresponding likelihoods assigned to the object. At the end of the process, an object was considered as a reliable H$\alpha$ emitter when the first probability exceeded the second emitter. Following this methodology, 46 out of the 202 candidates were finally selected as true H$\alpha$ emitters.

\subsection{H$\alpha$ line flux measurements}

The H$\alpha$ and [NII]6584 fluxes were derived from the individual pseudo-spectra of each H$\alpha$ emitter previously selected. Flux measurements were carried out by following the methodology in \cite{Sanchez-Portal2015}, which assumes infinitely thin lines and has the advantage of yielding non-contaminated fluxes for each line. Hence, a correction for the [NII] contribution in the case of the H$\alpha$ line flux is not necessary. Moreover, the availability of the fluxes of both lines allow us to diagnostic the AGN-host/star-forming separation, as described in Section \ref{LF_sec}. 

\subsection{Detection limits in OTELO}
\label{sec_simulaciones}
The detection limits and other biases of OTELO as a function of the emission-line parameters were obtained from extensive simulations of synthetic pseudo-spectra described below, based on the work in \cite{RamonPerez2017}. In our case, the number of false detections was kept under control by doing successive analysis of the emitting objects by independent collaborators (see Section \ref{sec_identification}). However, the number of missing objects had to be estimated. Therefore, the aim of the simulation we performed in this work was to find the detection limits of OTELO survey in terms of  EW, so as to see what kind of objects were escaping our detection when searching for H$\alpha$ emitters. To do so, we simulated an H$\alpha$ emission line in the form of a Gaussian and then convolved it to the spectral signature of OTELO to obtain its pseudo-spectrum. Three independent variables of the H$\alpha$ Gaussian were varied: its FWHM, its continuum flux density, and its amplitude. After ranging the grid of these variables a total of 500 independent simulations were performed. Each simulation is composed by one synthetic pseudo-spectrum for each node of the simulation grid in the FWHM-continuum-amplitude space. The synthetic pseudo-spectra are affected by random sky plus photon noise components scaled to the noise distribution of each resulting slice images of the OTELO tomography. Each spectrum was then convolved by the instrumental response of the tunable filter scan to obtain the simulated pseudo-spectra. We then checked whether the resulting pseudo-spectra were detected as emitters using the automatic algorithm for line detection mentioned above. \par 

\begin{figure}[!hbt]
\centering
\includegraphics[width=\linewidth]{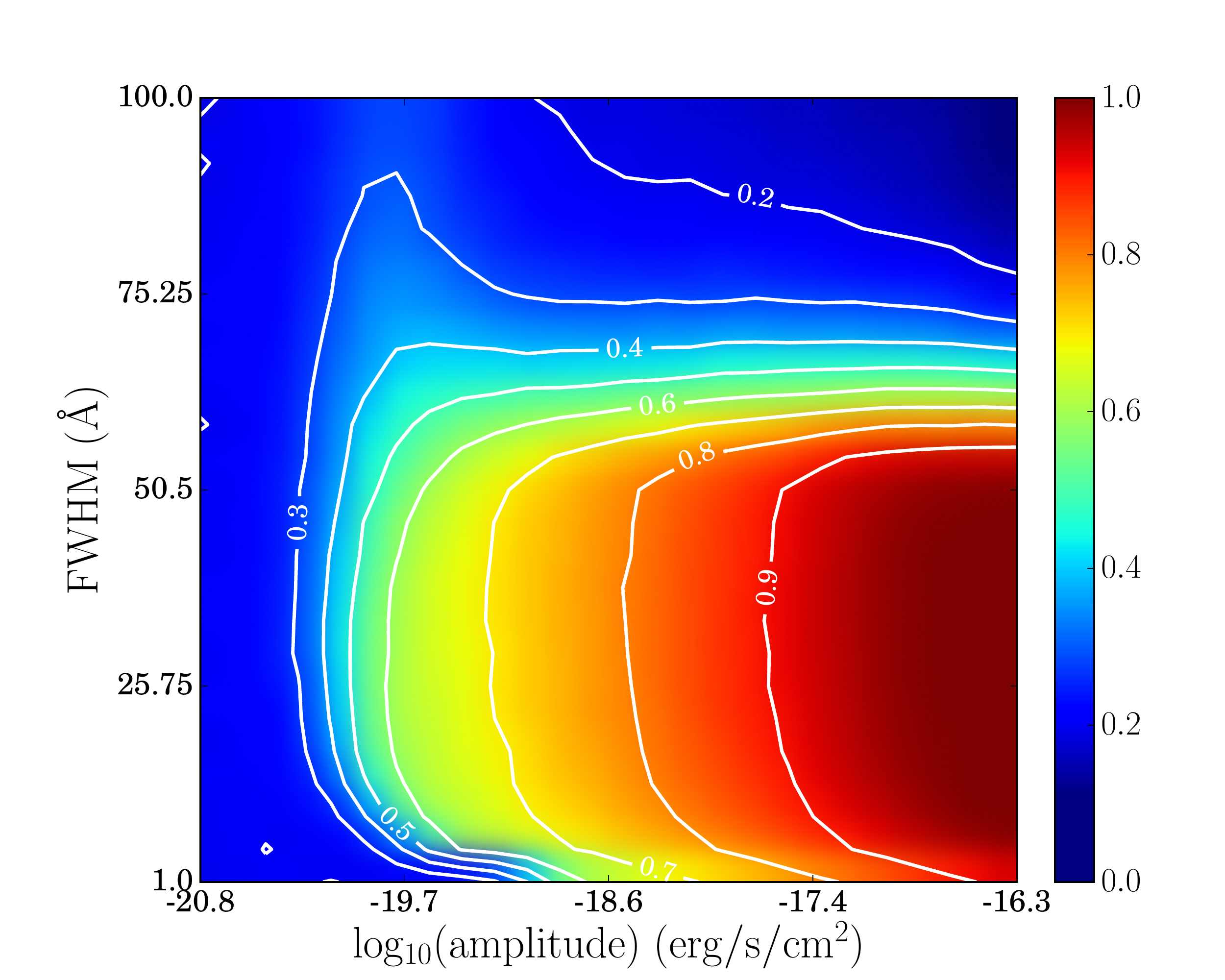}
\caption[]{Mean value of detection in the 500 runs of the simulation (0: never detected as emitter, 1: always detected as emitter), given the FWHM and the amplitude of the H$\alpha$ Gaussian. The continuum variable has been collapsed in the plane of the other two. White lines represent the detection probability contours.}
\label{simulationfig}
\end{figure}

As an example, Fig. \ref{simulationfig} shows the mean value of detection of the pseudo-spectra, given the original FWHM and amplitude of the H$\alpha$ emission line. There, a value of 0 means that the resulting pseudo-spectrum was never detected as an emitter, while a value of 1 means it was detected in the 500 runs of the simulation. We can see that for higher values of the amplitude and lower values of the FWHM, the detection fraction increases.

\begin{figure}[!hbt]
\centering
\includegraphics[width=\linewidth]{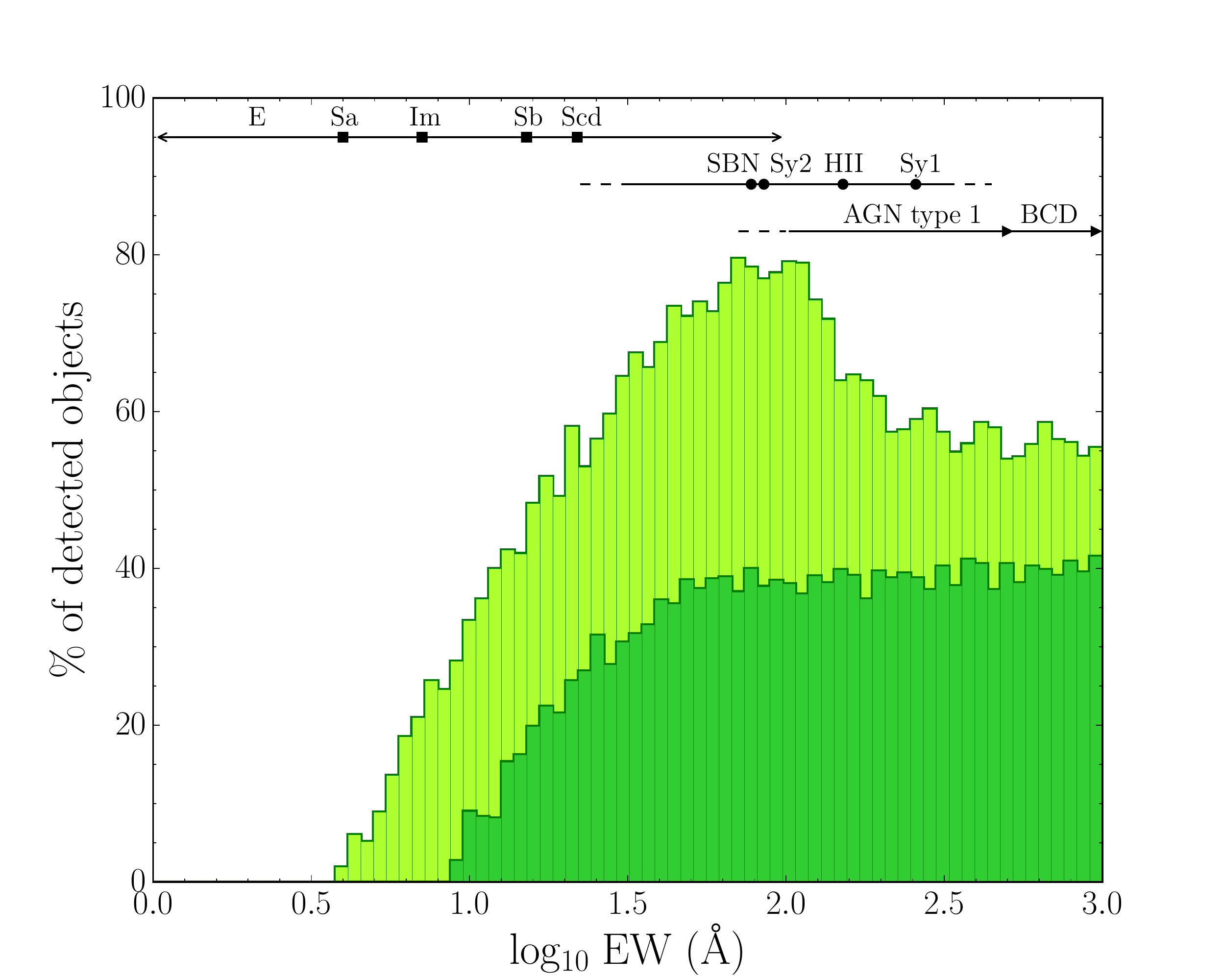}
\caption[]{Percentage of detected objects in the simulation depending on the EW of the input H$\alpha$ Gaussian line. Light green: $p\geq0.50$ probability threshold. Dark green: $p\geq 0.95$ probability threshold. Typical values of H$\alpha$ EWs for multiple astrophysical objects are also indicated. Black squares: median values of EW as in \cite{Gavazzi2006}. Black circles: mean values as in \cite{Gallego1997}. Black triangles: maximum values from \cite{Stern2012} and \cite{Gildepaz2003}.}
\label{simulationEW}
\end{figure}

In order to draw physical conclusions from the simulation, we computed for each experiment (i.e, for each combination of FWHM, continuum, and amplitude values) the  EW of the H$\alpha$ line. We determined the minimum detected EW for a given probability threshold $p$ (i.e. considering a level of detection of at least $p\times 100$\% in the whole simulation). For a probability threshold of $p\geq0.50$, for instance, the minimum detected EW is $\sim$5 \AA, while it is $\sim$10.5 \AA\ for a threshold of $p\geq0.95$. In Fig. \ref{simulationEW} we plot the percentage of detected objects in the simulation for those values of probability threshold as a function of the EW. Typical values of H$\alpha$ EWs for various astrophysical objects, taken from the works of \cite{Gavazzi2006}, \cite{Gallego1997}, \cite{Stern2012}, and \cite{Gildepaz2003}, are also indicated. \par

To summarise, the main results that can be drawn from the simulation about the detection limits of the H$\alpha$ line in OTELO survey are the following:  \par 

\begin{itemize}
\item[*] With a probability threshold of 50\% ($p\geq 0.50$) we should be able to detect 
\begin{itemize}
\item about 1 out of 5 elliptical galaxies with $\text{EW}=6$ \AA.
\item more than 75\% of spiral galaxies with $\text{EW} \geq 60$ \AA.
\item between 62\% and 79\%  of the galaxies with $100\leq \text{EW} \leq 200$ \AA.
\item between 55\% and 60\%  of the galaxies with $\text{EW} \geq 200$ \AA
.\end{itemize} 

\item[*] We reach a minimum EW of
\begin{itemize}
\item 5\ \AA$ $ with a probability of $p\geq 0.50$ for objects with a minimum flux density in the pseudo-continuum of $\sim 10^{-19}$ erg s$^{-1}$ cm$^{-2}$ \AA$^{-1}$.
\item 10.5\ \AA $ $ with a probability of $p\geq 0.95$ for objects with a minimum flux density in the pseudo-continuum of $\sim 10^{-18}$ erg s$^{-1}$ cm$^{-2}$ \AA$^{-1}$.
\end{itemize}
\end{itemize}

The results of this simulation are crucial in order to estimate the incompleteness of the sample of objects when deriving the H$\alpha$ LF (see section \ref{LF_sec}).

\section{Luminosity function at z$\sim$0.40} 
\label{LF_sec}

In this work, we obtained the H$\alpha$ LF at $z\sim 0.40$ and used this set of values to study the AGN contribution to the total luminosity at that redshift. The main advantage of OTELO, when compared to other surveys, is that emitters and AGN at $z\sim 0.40$ have been carefully inspected and selected. From the sample of H$\alpha$ emitters composed by 46 OTELO sources selected at $z\sim 0.40$, a total of 12 were optically classified as narrow- or broad-line AGN. The latter sources were segregated from the sample by using the EW$\alpha$n2 diagnostic \citep{CidFernandes2010}, in which the [OIII]/H$\beta$ ratio of the usual BPT diagram \citep{Baldwin1981} is replaced by the EW of H$\alpha$ at rest-frame. We adopted the prescription given by \cite{Stasinska2006} to separate pure star-forming galaxies from composite/AGN hosts in the sample. More details about this particular topic and the AGN demographics in OTELO are given in OTELO-III. This implies that the AGN contribution in each luminosity interval is exactly known and not only estimated, as in other works (see  e.g. \citealt{Shioya2008} or \citealt{Sobral2013}). \par

\subsection{Completeness correction}
\label{completeness}

\begin{figure}[!htb]
\centering
\includegraphics[width=\linewidth]{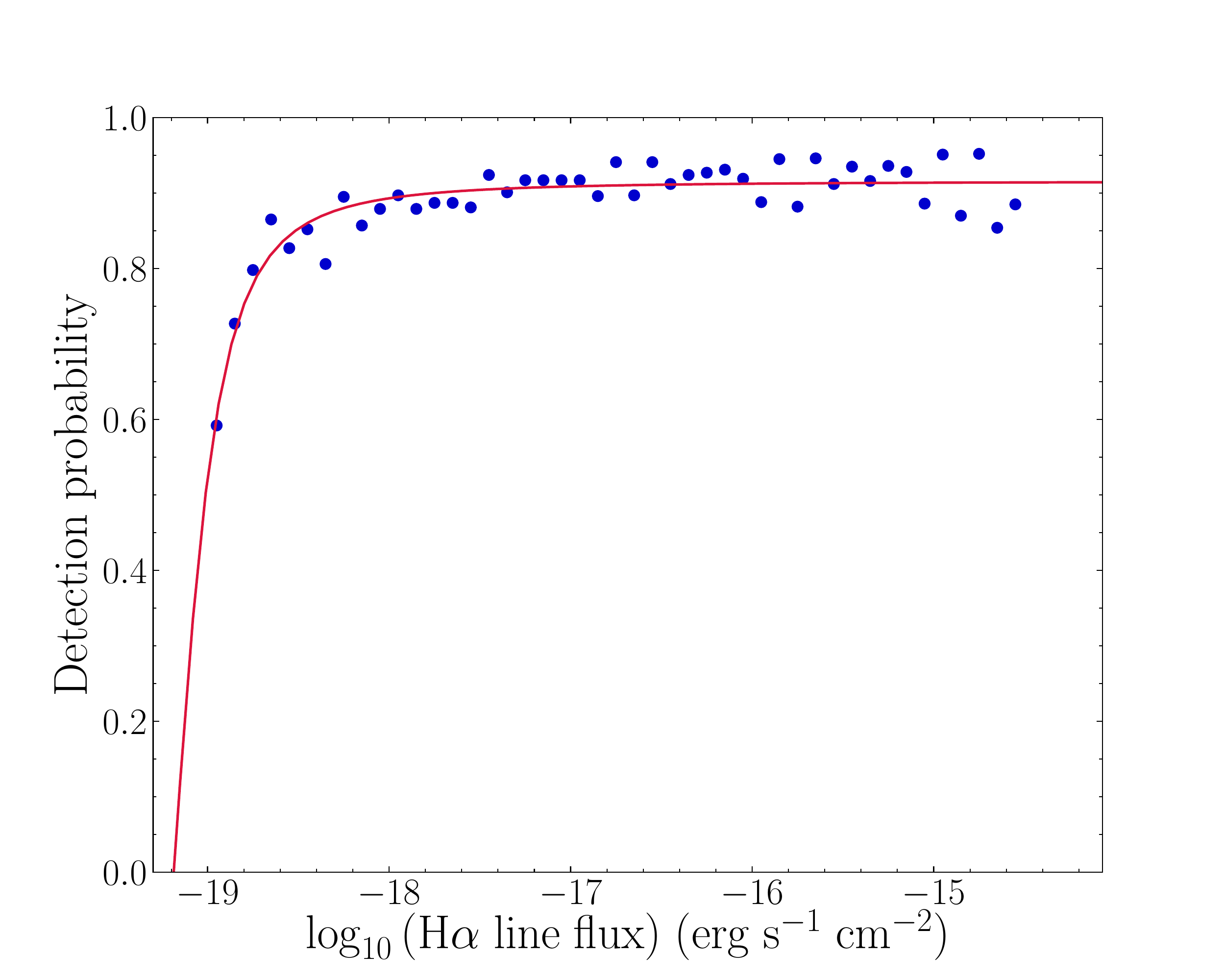}
\caption[Mean detection probability as a function of H$\alpha$ line flux]{Mean detection probability as a function of H$\alpha$ line flux (blue dots) for simulated objects having $p\geq 0.50$ (see Section \ref{sec_simulaciones}). The red line indicates the best least-squares fitting of a sigmoid function (see description in the text).}
\label{probdet}
\end{figure}

One of the main difficulties when deriving the LF is to estimate (and correct) the incompleteness of the sample of objects. In this work, we take advantage of the simulations described in section \ref{sec_simulaciones}, which evaluated the detection probability of an H$\alpha$ emission line at $z\sim 0.40$ in OTELO. We calculated the mean detection probability as a function of the line flux, for the simulated objects having $p\geq0.50$, which is the completeness cut that we adopted. The result is shown in Fig. \ref{probdet}. We then fitted the data to a sigmoid function such as $d=a F_{\rm l} / \sqrt{c-F_{\rm l}^{2}}$, where $F_{\rm l}=\log(f_{\rm l})-b$, $f_{\rm l}$ is the line flux. The values obtained from a least-squares fitting using the Levenberg-Marquardt algorithm implementation in {\tt IDL} were $a=0.916\pm0.005$, $b=19.187\pm0.047,$ and $c=0.072\pm0.023$. This function was used as our LF completeness correction.

\subsection{Volume sampled and cosmic variance}
\label{variance}
According to the specific characteristics of the OTELO TF-tomography in the spectral dimension and the effective field of view (FoV) sampled, we estimated the comoving volume limited by the H$\alpha$ line at $0.37\leq z_{spec}\leq 0.42$, amounting $1.4\,\times\,10^{3}$ Mpc$^3$. Because of the relatively small volume of Universe probed for this emission line, the effects of the cosmic variance are not negligible; see \citealt{Stroe2015} to better understand the impact of this effect on the LF determination. The mean comoving number density of all the OTELO H$\alpha$ sources is $0.033$ Mpc$^{-3}$. Following \cite{Somerville2004} we estimated a galaxy bias $b$ of $\sim0.77$ for this number density, and a variance of DM, $\sigma_{\rm DM} = 0.95$ for such a volume. Accordingly, a general cosmic variance is $\sigma_{\rm v} = b\,\sigma_{\rm DM} \simeq 0.73$ in this science case.\par

\subsection{H$\alpha$ luminosity function}
\label{HaLF}

The LF calculation starts by computing the number $\Phi$ of galaxies per unit volume and per unit of H$\alpha$ luminosity, in bins of $\Delta[\log L(\text{H}\alpha)] = 0.5$. We used the $V_{\rm max}$ method (\citealt{Schmidt1968}; see, for instance, Eq. 2 in \citealt{Bongiovanni2005} or Eq. 8 in \citealt{Hayashi2018}) with the following adapted expression:

\begin{center}
\begin{equation}
\Phi[\log_{10} L(\text{H}\alpha)] = \kappa \frac{4\pi}{\Omega} \sum_{\rm i} \frac{1}{d_{\rm i}},
\end{equation}
\end{center}

\noindent 
which takes into account the volume of Universe surveyed and the completeness bias in our survey:  $d_{\rm i}$ is the detection probability for the i-$th$ galaxy, $\Omega$ is the observed solid angle ($\sim 4.7 \times 10^{-6}$ str), and $\kappa$ is a normalisation factor proportional to $V_{\rm max}^{-1}$ ($\kappa=V_{\rm max}^{-1} \times \Delta[\log L(\text{H}\alpha)]^{-1}$); the maximum comoving volume is limited by the H$\alpha$ line in OTELO spectral range.\par 

The H$\alpha$ sample was corrected from completeness following the function shown in Fig. \ref{probdet} and also from dust attenuation, using the reddening value given by the best spectral energy distribution fit from {\it LePhare} for each galaxy. We used the empirical relation by \cite{Ly2012}, which connects the dust reddening $E(B-V)$ and the extinction $A_{\text{H}\alpha}$ for galaxies at $z\sim 0.50$: $A_{\text{H}\alpha}=1.85 \times E(B-V) \times k(\text{H}\alpha) $, where $k(\text{H}\alpha)=3.33$ and the proportionality factor comes from the ratio between stellar and nebular reddening as follows: $E(B-V)_{\rm gas}=1.85 \times E(B-V)$ \citep{Calzetti2000}. \par

\begin{figure*}[!hbt]
\centering
\includegraphics[width=0.8\textwidth]{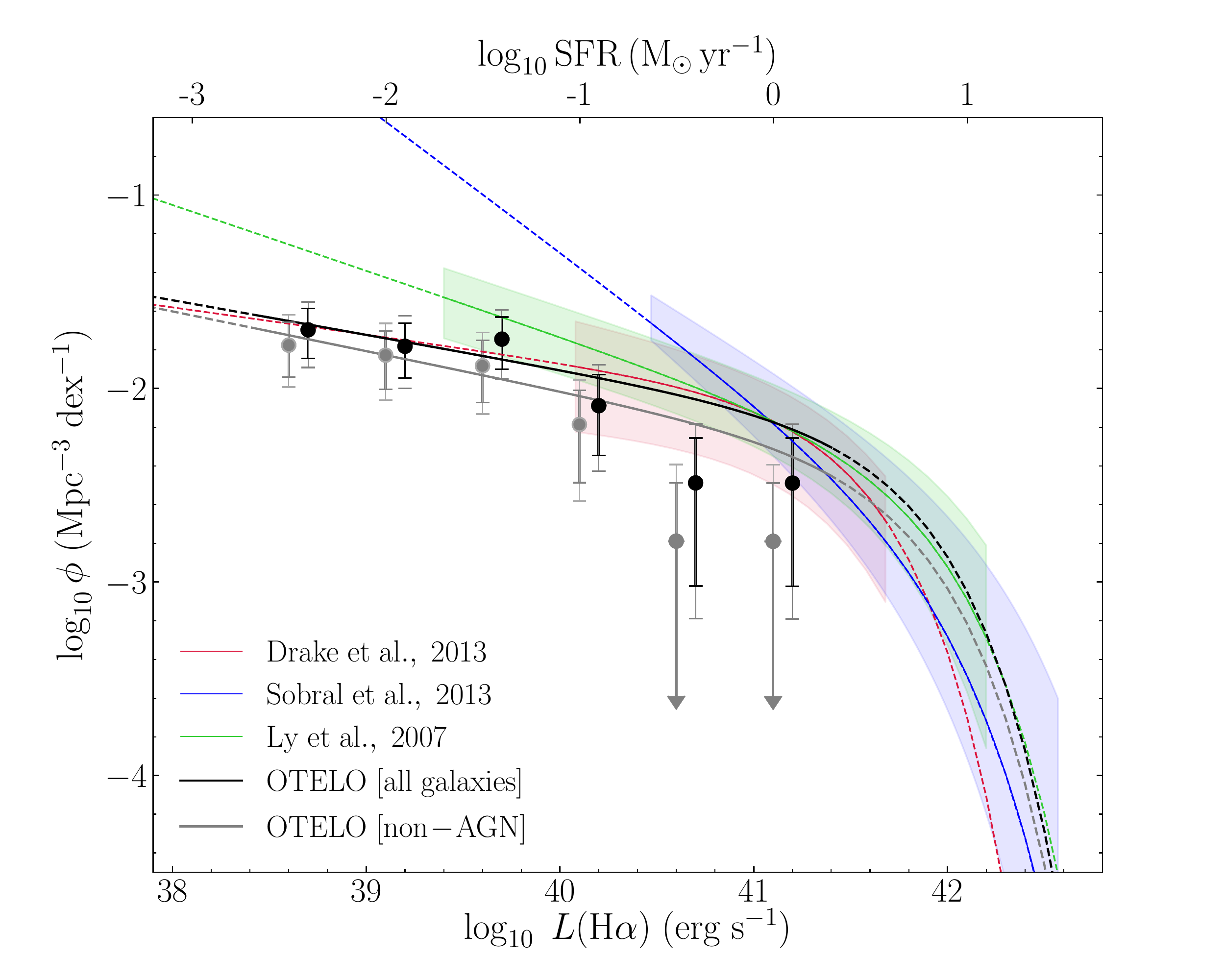}
\caption[Completeness and dust extinction corrected H$\alpha$ luminosity function at $z\sim 0.40$]{Completeness and dust extinction-corrected H$\alpha$ LFs at $z\sim 0.40$. The black line corresponds to the LF fitting for the whole sample of OTELO emitters at that redshift, which are represented with black dots, while the grey line is the LF fitting for the non-AGN population (grey circles). For the sake of clarity, grey circles are displaced -0.1 dex in luminosity with respect to black circles. Shorter error bars are Poissonian and the larger error bars have the cosmic variance uncertainties evenly added in quadrature (see text for details). The H$\alpha$ LFs of \cite{Drake2013}, \cite{Sobral2013}, and \cite{Ly2007} are plotted in red, blue, and green, respectively. In each case, the solid line represents their sampled luminosity interval, while the dashed line is the extrapolation of their Schechter function fit. The coloured regions are the maximum envelopes that enclose binned data and their corresponding errors as reported by the cited authors.}
\label{LF}
\end{figure*}

Using the aforementioned expressions, a completeness and dust extinction-corrected H$\alpha$ LF at $z\sim0.40$ was obtained (see Table \ref{points_LF}) for the OTELO complete data set (46 objects) and for the OTELO star-forming galaxies sample (34 objects). A Schechter function \citep{Schechter1976} was then fitted to our data points. Accounting for our low number statistics for the determination of the LFs, and following the data weighting analysis in \cite{Comparat2015}, we adopted the ratio of the number of galaxies within each luminosity bin to the total number in each subsample as weighting scheme in our fitting. The Schechter parameters of the OTELO LFs for each subsample are given in Fig. \ref{LF}, along with those obtained by previous works for H$\alpha$ emitters at the same redshift. Main error sources in our LFs are those of statistical nature (i.e. obeying a Poissonian process) and those derived from the cosmic variance effects. The contribution of these uncertainties were summed in quadrature and dominate by far over those that come from the line flux calculations and those corresponding to the detection probability function represented in Figure \ref{probdet}. For this reason the latter sources of error were dismissed in our LF estimations. \par

The characteristics of OTELO survey makes it very competitive in terms of depth and recovery of the LF faint end. As can be seen in Fig. \ref{LF}, OTELO data reach luminosities almost 1 dex fainter than the faintest limit of the deepest survey \citep{Ly2007}. However, owing to its small angular coverage and number statistics, the  LF bright end of OTELO is poorly sampled. That is why we decided to fix the $L^\ast$ parameter and take the mean value $\bar{L^*}$ from the LFs of \cite{Drake2013}, \cite{Sobral2013} and  \cite{Ly2007}: $\log_{10}\bar{L^*}$ (erg s$^{-1}$) = 41.85. The dispersion in $\bar{L^*}$ from data in those works is small ($\sigma_{\bar{L^*}} = 0.16$), reinforcing our assumption. Besides, the strength of OTELO resides in its depth, and hence in determining the slope at the faint end, $\alpha$. This parameter and $\phi^*$ were obtained by fitting the Schechter function to our data points. Their values, as well as those obtained by the already mentioned works, are shown in Table \ref{Table_LF}.\par
 
 \begin{table}[ht]
\vspace*{2mm} 
\caption[]{H$\alpha$ LFs (corrected by dust extinction and completeness) at $z\sim0.40$ for the OTELO survey, considering all the objects in the sample (46) and the non-AGN only (34). Errors are subject to Poissonian statistics only.}
\vspace*{-2mm}       
\label{points_LF}      
\centering        \small                          
\begin{center}
\addtolength{\tabcolsep}{-1.5pt}
\begin{tabular}{ccc}   
\hline\\[-2pt]
 H$\alpha$ sample & $\log_{10} L$(H$\alpha$) (erg s$^{-1}$) & $\log_{10}\phi$ (Mpc$^{-3}$dex$^{-1}$)  \\[3pt] \hline \\[1pt]
 OTELO [all galaxies] & 38.70 & -1.697$^{+0.11}_{-0.15}$ \\
\\[0.05em]
                     & 39.20 & -1.782$^{+0.12}_{-0.17}$ \\
\\[0.05em]
                     & 39.70 & -1.745$^{+0.12}_{-0.16}$ \\
\\[0.05em]
                     & 40.20 & -2.089$^{+0.16}_{-0.26}$ \\
\\[0.05em]
                     & 40.70 & -2.488$^{+0.23}_{-0.53}$ \\
\\[0.05em]
                     & 41.20 & -2.489$^{+0.23}_{-0.53}$ \\
\\[0.05em]
& &    \\[0.3em]
OTELO [non-AGN] & 38.70 & -1.776$^{+0.12}_{-0.17}$ \\
\\[0.05em]
                & 39.20 & -1.828$^{+0.13}_{-0.18}$ \\
\\[0.05em]
                & 39.70 & -1.883$^{+0.13}_{-0.19}$ \\
\\[0.05em]
                & 40.20 & -2.186$^{+0.18}_{-0.30}$ \\
\\[0.05em]
                & 40.70 & -2.789$^{+0.30}_{-2.40}$ \\
\\[0.05em]
                & 41.20 & -2.790$^{+0.30}_{-2.40}$ \\
\\[0.1em]
 \hline                                             
\end{tabular}\end{center}
\end{table}

\begin{table*}[ht]
\vspace*{2mm} 
\caption[Schechter parameters of the H$\alpha$ luminosity functions at $z\sim0.40$] Schechter parameters of the dust-corrected H$\alpha$ LFs for the OTELO survey and earlier works, all at $z\sim 0.40$.
\vspace*{2mm}       
\label{Table_LF}      
\centering        \small                          
\begin{center}
\addtolength{\tabcolsep}{-2.0pt}
\begin{tabular}{ccccccccc}   
\hline\\
Data set & Number of & Volume & $\log_{10} L_{\rm lim}$ & $\log_{10} \phi^*$ & $\log_{10} L^*$ & $\alpha$ & $\log_{10}\cal{L}$ & $\rho_{\rm SFR}$ \\
\\
 & sources ($N$) & (10$^3$ Mpc$^{3}$) & (erg s$^{-1}$) & (Mpc$^{-3}$ dex$^{-1}$) & (erg s$^{-1}$) &   & (erg s$^{-1}$ Mpc$^{-3}$) & (M$_{\odot}$\ yr$^{-1}$\ Mpc$^{-3}$)  \\
\\
\hline     \\[0.5em]
\cite{Ly2007} &  391 & 4.71 & 39.6 & -2.75$\pm$0.16 & 41.93$\pm$0.19 & -1.34$\pm$0.06 & 39.31$\pm$0.08 & 0.016$\pm0.003$ \\
\\[0.4em]
\cite{Drake2013} &  271 & 29.5 & 40 & -2.44$^{+0.14}_{-0.17}$ & 41.55$^{+0.13}_{-0.11}$ & -1.14$^{+0.14}_{-0.13}$ & 39.15$\pm$0.02 & $0.0113\pm0.0005$  \\
\\[0.4em]
\cite{Sobral2013} & 797 & 88 & 40.5 & -3.12$^{+0.10}_{-0.34}$ & 41.95$^{+0.47}_{-0.12}$ & -1.75$^{+0.12}_{-0.08}$ & 39.55$\pm$0.22 & $0.03\pm0.01$ \\  
\\[0.4em]
OTELO [all galaxies] & 42 ($<L*$)& 1.4 & 38.7 & -2.59$\pm$0.22 & 41.85 (fixed) & -1.18$\pm$0.08 & 39.31$\pm$0.18 & --\\
\\[0.4em]
OTELO [non-AGN] &33 ($<L*$)& \texttt{"} & 38.7 &-2.75$\pm$0.19 & 41.85 (fixed) & -1.21$\pm$0.07 & 39.17$\pm$0.16 & $0.012\pm0.005$  \\
\\[0.4em]
\hline
\end{tabular}\end{center}
\end{table*}

The OTELO LFs allow us to extend the luminosity range almost 1 dex fainter than previous works such as \cite{Ly2007}. The LF of \cite{Sobral2013} seems to deviate from the rest at faint luminosities, overestimating the number of low-luminosity objects. However, given the luminosity limits of their survey, this should not be surprising. The work of \citeauthor{Sobral2013} is, nevertheless, a good reference for the bright end of the LF, taking into consideration the relatively large comoving volume by them explored (see Table \ref{Table_LF}). It is also worth noticing the potential of OTELO survey to recover and extend the LF, in spite of the small number of objects under study in our sample. As a matter of fact, the samples from works by \cite{Drake2013}, \cite{Sobral2013}, and  \cite{Ly2007} have 6.5, 19, and 9 times more galaxies than our own, respectively, and with $\sim 4$ to 60 times the comoving volume explored by OTELO at this redshift. Even so, at the faint end, OTELO's H$\alpha$ LF at $z = 0.4$ is in agreement with the extrapolation of the SFR function of \cite{Drake2013}, with a shallower slope compared to those provided by \cite{Ly2007} and \cite{Sobral2013}. \par 

The 34 objects in our sample of H$\alpha$ emitters that were not selected as AGN are presumably star-forming galaxies. By studying their LF, the SFR can be derived following the standard calibration of \cite{Kennicutt1998}, which assumes a Salpeter initial mass function (IMF) with masses between 0.1 and 100 M$_{\odot}$:

\begin{center}
\begin{equation}
\text{SFR} (\text{M}_\odot\text{ yr}^{-1}) = 7.9\times 10^{-42} \text{ }L\,(\text{H}\alpha)\,({\rm erg}\ {\rm s^{-1}}).
\end{equation}
\end{center}

Integrating this LF we obtain a value of $\log_{10}{\cal{L}}(\text{H}\alpha)= 39.17\pm 0.16$, which translates into a SFR density of $\rho_{\rm SFR}=0.012\pm0.005\ {\rm \text{M}_{\odot}\ yr^{-1}\ Mpc^{-3}}$ at $z\sim 0.40$. This value is in agreement with that provided by \cite{Drake2013}, but it is also consistent with other obtained by previous works (see last column in Table \ref{Table_LF}). It is worthy of note that this estimation does not include the contribution of the AGN hosts to the star formation budget in the comoving volume explored. Our SFR density estimation is closer to that of \cite{Ly2007}  and \cite{Drake2013}, while the value obtained by \cite{Sobral2013} departs from the rest and probably tends to overestimate the SFR for the aforementioned reasons. \par

\subsection{Role of AGN hosts}
\label{agncont}
\begin{figure}[h]
\centering
\includegraphics[width=\linewidth]{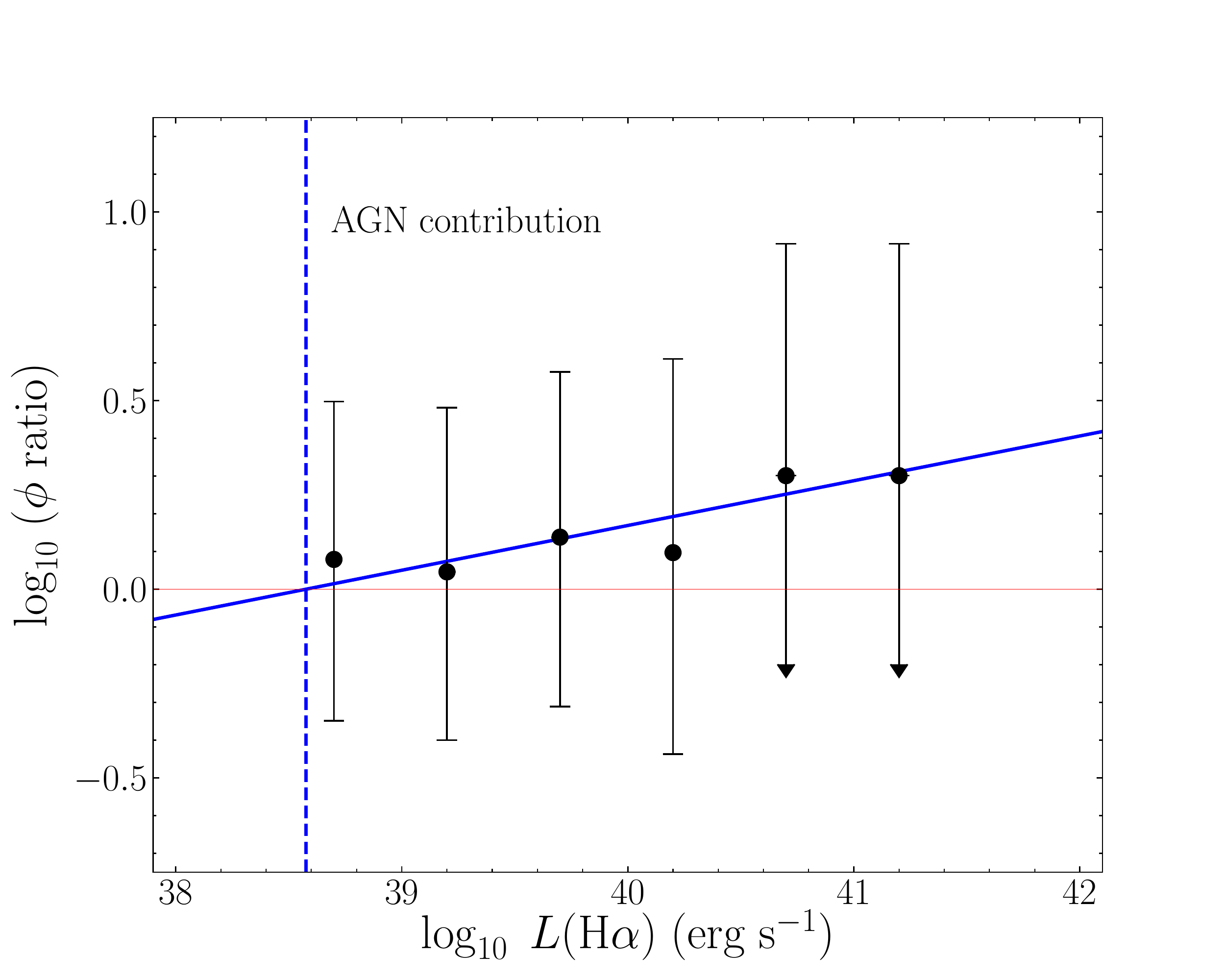}
\caption[Ratio between the OTELO H$\alpha$ LF of the total sample of emitters at $z\sim0.40$ and that of the star-forming galaxies only]{Ratio between the OTELO H$\alpha$ LF of the total sample of emitters at $z\sim0.40$ and that of the star-forming galaxies only. Bars are the sum in quadrature of the Poissonian errors of both LF, as given in Table \ref{points_LF}. The blue solid line represents a error-weighted linear fitting to the ratio values. The red horizontal line corresponds to a ratio equals unity. The blue dashed vertical line represents the luminosity above which the AGN contribution becomes relevant.}
\label{LFratio}
\end{figure}

The AGN contribution to the total H$\alpha$ luminosity at $z\sim0.40$ is estimated by comparing the OTELO LF obtained for the complete sample of emitters and that obtained for the sample of star-forming galaxies only, both represented in Fig. \ref{LF}. The ratio of the LFs given in Table \ref{points_LF} is represented in Figure \ref{LFratio}), and it can be fitted with a linear function with the following parameters:

\begin{align*}
\log_{10}\left(\frac{\phi_{\rm \,[all\ galaxies]}}{\phi_{\rm \,[non-AGN]}}\right)  &= a\times\log_{10} L\text{(H}\alpha\text{)}+b  \numberthis \label{LF_diff_eq}, \\
a &= 0.119 \pm  0.033   \nonumber \\
b &= -4.577 \pm 1.335. \nonumber
\end{align*}

The contribution of AGN hosts in the H$\alpha$ luminosity range explored by OTELO at z=0.4 tends to increase monotonically up to a factor $\sim1.5$ starting from $\log_{10}$L(H$\alpha$) (erg s$^{-1}$) $\gtrsim$ 38.6. Therefore, it would not be expected to find galaxies hosting AGN at $z\sim 0.4$ below that luminosity threshold. This trend is in agreement with the results obtained by \cite{Sobral2016} and \cite{Matthee2017}. In this sense, \cite{Gunawardhana2013} found that the highest AGN contribution in the H$\alpha$ LFs occurs at $0.17 < z < 0.24$, even though this effect is almost negligible at $z<0.1$. \par

Given the small number of objects in our H$\alpha$ sample for the construction of the LFs, the above results about the AGN contribution should be taken with caution. However, it is worth noticing that the general procedure is addressed to estimate the fraction of AGN as a percentage over the total population, and correct the sample of galaxies accordingly to build the LF. The OTELO survey, in contrast, has allowed us to obtain a complete sample of H$\alpha$ emitters, for which the AGN fraction is accurately known at each luminosity interval, despite the small comoving volume sampled.\par

\section{Conclusions}\label{conclusions}

This work has focussed on the OTELO survey, an emission-line object survey using the red tunable filters of the OSIRIS instrument at the Gran Telescopio Canarias \citep{OTELO1}. Following the selection of emitting objects described in \cite{OTELO1}, H$\alpha$ emitters candidates at $z\sim0.40$ have been identified and a sample of 46 emitters has been selected.\par 

Then, an H$\alpha$ LF at $z\sim0.40$ has been constructed, using a completeness correction derived from a simulation of the detection limits in OTELO and a dust extinction correction from the photo-$z$ estimation \citep{OTELO1}. Using a mean value of $\log_{10}\bar{L^*}$ (erg s$^{-1}$) = 41.85 ($\sigma_{\bar{L^*}} = 0.16$) calculated from data of previous works, we fitted our data points to a Schechter function with the following parameters for the whole sample of H$\alpha$ emitters: $\log_{10}\phi^* = -2.59\pm0.22$ and  $\alpha= -1.18\pm0.08$. Using the EW$\alpha$n2 diagnostic \citep{CidFernandes2010}, we removed the AGN from the previous sample and obtained a second LF for non-AGN only with the following parameters: $\log_{10}\phi^* = -2.75\pm0.19$ and  $\alpha= -1.21\pm0.07$.\par 

When compared to previous works, our LFs extend the faint end almost 1 dex, reaching observed H$\alpha$ luminosities as low as  $\log_{10}L_{\rm lim} ({\rm erg\ s}^{-1})$ = 38.5 (equivalent to a SFR of $\sim0.002\, \text{M}_\odot\text{ yr}^{-1})$, and therefore constricting the faint-end slope at such luminosity regime.\par

The AGN contribution to the total H$\alpha$ luminosity has been estimated. We find that no AGN should be expected below a luminosity of $\log_{10}L_{\rm lim} ({\rm erg\ s^{-1}})$ = 38.6. Above this value, the AGN contribution grows with the luminosity, in agreement with previous works. Again, given our small sample of AGN at $z\sim0.40$, a study with more statistical significance should be conducted to confirm this result. Discarding AGN hosts contribution to the star formation, we obtained an integrated H$\alpha$ luminosity from star-forming galaxies at $z=0.40$ of $\log_{10}\cal{L}\, ({\rm erg\ s^{-1}})$ = 39.17$\pm$0.16, yielding a SFR density of $\rho_{\rm SFR}=0.012\pm0.005\ {\rm \text{M}_{\odot}\ yr^{-1}\ Mpc^{-3}}$.

\bigskip
\bigskip

\begin{acknowledgements} This  work  was  supported  by  the  Spanish  Ministry  of  Economy  and  Competitiveness  (MINECO) under  the  grants 
AYA2013\,-\,46724\,-\,P,
AYA2014\,-\,58861\,-\,C3\,-\,1\,-\,P,
AYA2014\,-\,58861\,-\,C3\,-\,2\,-\,P,
AYA2014\,-\,58861\,-\,C3\,-\,3\,-\,P,
AYA2016\,-\,75808\,-\,R,
AYA2016\,-\,75931\,-\,C2\,-\,2\,-\,P,
AYA2017\,-\,88007\,-\,C3\,-\,1\,-\,P, and
AYA2017\,-\,88007\,-\,C3\,-\,2\,-\,P.
Based on observations made with the Gran Telescopio Canarias installed in the Spanish Observatorio del Roque de los Muchachos of the Instituto de Astrof\'isica de Canarias, in the island of La Palma. The authors thanks the anonymous referee for her/his feedback and suggestions.
\end{acknowledgements}

\bibliographystyle{aa}
\bibliography{otelo_HalF_final}

\end{document}